\documentclass[preprint,aps]{revtex4}
\usepackage{graphicx}
\usepackage{booktabs}
\usepackage[usenames, dvipsnames]{color}
\usepackage{multirow}
\usepackage{eurosym}
\usepackage{setspace}
\usepackage{rotating}
\usepackage[ pdftex, plainpages = false, pdfpagelabels, 
                 pdfpagelayout = useoutlines,
                 bookmarks,
                 bookmarksopen = true,
                 bookmarksnumbered = true,
                 breaklinks = true,
                 linktocpage=all,
                 pagebackref=false,
                 colorlinks = true,
                 linkcolor = BrickRed,
                 urlcolor  = blue,
                 citecolor = BrickRed,
                 anchorcolor = green,
                 hyperindex = true,
                 hyperfigures
                 ]{hyperref}

\begin{document}

\title{\href{http://www.necsi.edu/research/social/egypt}{Political Stability and Military Intervention in Egypt$^\dagger$
}}
\date{July 8, 2013}  
\author{Casey Friedman, Dominic K. Albino, and \href{http://necsi.edu/faculty/bar-yam.html}{Yaneer Bar-Yam}}
\affiliation{\href{http://www.necsi.edu}{New England Complex Systems Institute} \\ 
238 Main St. S319 Cambridge MA 02142, USA } 

\begin{abstract}
Policy choices in the wake of recent mass protests in Egypt will determine the likelihood of civil war in the short run and the prospects for democracy in the long run. Economic conditions can be improved by international action to reduce grain-based biofuel production and finance employment generation. Creating the conditions for stable democracy requires accepting power-sharing mechanisms in which the military will have an important role to ensure stability. 
 \\ \\\\\ \\\\\\\\\\\\\\\\\\\\\\\
$\dagger$ 
\footnotesize Prepared for presentation to the Chairman's Action Group, Office of the Chairman of the Joint Chiefs of Staff, Pentagon on Wednesday, July 3, 2013.
\end{abstract}

\maketitle

Turmoil in Egypt has returned two and a half years after the fall of Mubarak in 2011. Risks of an outright civil war are high.
The military has deposed the elected president Mohammed Morsi in response to the largest street protests in Egyptian history, which involved millions of people. Unrest has not been halted.
These events are consistent with expectations from our published research. 

Recent NECSI work shows that revolutions are more likely to result in autocracy than in stable democracy---even when democracy is the explicit goal of those seeking change \cite{revolutions}. Revolutions do not allow adequate time for stable democratic institutions to arise; the pressing economic and social demands on the government are high and unrealistic. 
Spiraling discontent poses a continuing and worsening risk of anarchy unless an autocratic system re-establishes order. At the same time, the creation of a viable democratic system requires sufficient stability to develop democratic institutions that can successfully achieve legitimacy in the eyes of the public by mediating between competing interests and meeting economic needs. 

On the surface, it seems that the Egyptian military's removal of Morsi has undermined democracy. However, political stability is essential to the creation of a successful democratic regime. In the post-revolutionary context, there is no real trade-off between democracy and peace, because 
stability is necessary for democracy to emerge.

If there had been no military action to remove Morsi, the pattern of recent protests suggests that millions of people would have stayed in the streets, and the state would have been unable to impose adequate order. Police forces abandoned the effort to impose order. The military is the only entity with the capacity to avert violence on a substantial scale.

Thus, Egypt will have the best chances of positive outcomes for both preventing civil war and increasing the prospects of stable democracy if the military's intervention can preserve order and create a stable transition process to a new government. Whether the current military-led transition will establish sufficient political stability remains to be seen. Moreover, action must be taken to address the underlying economic problems. 

NECSI work shows that among the key drivers of unrest are high food prices that foster desperation \cite{food_crises,food_prices,Feb_update,July_update,food_for_fuel,food_faq,yemen}.
 The current global measure of food prices reported by the UN Food and Agricultural Organization (FAO) is at 215, double its value ten years ago, and just above the threshold at which NECSI research shows widespread violence 
takes place \cite{food_crises}.  Food prices are largely set on international markets, where Egyptian government policies have no influence, and the burden of high prices must be borne either by the government---in the form of expanded food subsidies---or by the Egyptian public. 

Egypt's economic weaknesses extend beyond food prices. 
There is absence of opportunity and employment for many, especially youth. Economic growth has stagnated since the revolution in 2011 due to social disruption and consequent loss of economic activity including tourism and foreign investment \cite{econnytimes, econbloomberg}. Failures of government services have occurred as a result of a fiscal crisis and the evaporation of foreign exchange reserves.

Solving these diverse and interrelated economic problems requires a multi-pronged approach.
First, the US must take policy action to reduce corn-to-ethanol conversion, leading to lower global food prices and dramatically reducing the immediate pressure on poor populations. A bill to eliminate the mandate for ethanol production is pending in Congress \cite{goodlatte}, but further action may be required to actually reduce ethanol production.

Given the current deepening financial and economic crisis in Egypt, political stability also depends on securing financing for the government budget and reducing unemployment. 
In light of its fragile political climate, Egypt can only accomplish these goals with a major aid package that can overcome the immediate fiscal crisis and reduce the economic desperation.

Negotiations with the International Monetary Fund have carried on for two years without conclusion of a \$4.8 billion loan deal because the Fund insists on austerity measures that could not be made without prompting violence and chaos. In particular, it is not feasible for Egypt to reduce food subsidies while high food prices and unemployment prevail. Saudi Arabia, Qatar, and Libya have been the main financial backers of the Egyptian government since the 2011 revolution. These governments understand that collapse of the Egyptian economy would be a cataclysm for the region.

As a more comprehensive economic solution, we have advocated a major public works program, which regional oil-rich countries can support financially, and to which outside groups such as the United States Army Corps of Engineers can contribute technical assistance \cite{usace}. Such a program is an appropriate response to the current conditions and the threat of widespread social disorder that would result from inaction.

As an immediate, practical measure, the United States should maintain the flow of economic assistance to Egypt. US law requires the suspension of aid to any country in which a coup d'\'{e}tat deposes an elected government. However, it is better to consider the social unrest and military intervention as part of the process of the government transition that began with the fall of Hosni Mubarak. Maintaining the flow of aid will not repair the economic situation in Egypt, but it will help avoid further deterioration. 

In addition to economic relief and recovery, Egypt needs stable political structures. The major political failure of the prior government was its unwillingness to protect the interests of electoral minorities, making recurring conflict likely. Democracy depends on the existence of a government acceptable to a broader segment of the population than a simple majority. It must not be rejected by any minority large enough to cause social disorder.  Ideologies and demands that are wholly excluded from governance undermine stability. When the populace has opinions that are narrowly distributed around a well-defined mean, democratic governance is readily achieved. When there are broad distributions of opinion relative to accepted levels of tolerance, or well separated groups, the ability to achieve legitimacy for governance is undermined. Decreasing tolerance and increasing divergence of groups undermines the ability of conventional democratic governance.

Democracy as many American commentators understand it is incompatible with the Egypt of today. Islamists and secular Egyptians have markedly different aims for the country's future. At the time of the 2011 revolution, the Muslim Brotherhood portrayed itself as occupying this middle ground, but subsequent events have belied this public image. Consequently, neither Islamists nor moderates will acknowledge the legitimacy of governance by the other. A stable Egyptian political system must therefore be structured, formally or informally, to ensure that each group with the power to undermine social order has a strong motivation to preserve it. The Lebanese ``troika'' arrangement -- which requires a Maronite president, Shi'ite speaker of parliament, and Sunni prime minister -- offers a useful example of a potential power-sharing system. In Egypt, the relevant groups are the Islamists, the moderates (represented by the National Salvation Front), and the military. It is likely to be helpful to incorporate the military in government as balancing  three groups can be more stable than two groups, though any balance requires a willingness to compromise. As in any democracy the possibility of sub-factions being accommodated through compromise must be present.

Important goals for international engagement in Egypt in light of the recent mass protests and ouster of the Morsi government are, first, to avoid a major civil war in the short run and, second, promote a stable government over the longer term. If economic and political challenges are not overcome, there is a very strong possibility of protracted, large-scale violence like that now seen in Syria. Unlike Syria, Egypt is not divided into geographically distinct enclaves, with the implication that violence could be even more destructive and widespread in Egypt.

We thank Karla Z. Bertrand, Maya Bialik and Roozbeh Daneshvar for helpful comments.

\newpage
\bibliographystyle{Science}

\end{document}